\documentclass[conference]{IEEEtran}
\IEEEoverridecommandlockouts
\usepackage{cite}
\usepackage{times} 
\usepackage{amsmath,amssymb,amsfonts}
\usepackage{algorithm}
\usepackage{graphicx}
\usepackage{textcomp}
\usepackage{amsmath, amssymb, amsthm}
\usepackage{bm}
\usepackage{enumitem}
\usepackage{algpseudocode}
\usepackage{hyperref}
\usepackage{tikz}          
\usepackage{xcolor}
\usepackage{setspace}      

\newcommand{\titlefont}{\fontsize{24pt}{30pt}\selectfont} 
\newcommand{\authorfont}{\fontsize{10pt}{12pt}\selectfont} 
\newcommand{\affilfont}{\fontsize{10pt}{12pt}\selectfont}  
\newcommand{\abstractfont}{\fontsize{10pt}{10pt}\selectfont} 
\newcommand{\keywordfont}{\fontsize{10pt}{10pt}\selectfont}  

\begin{document}

\title{\titlefont An ACO–MPC Framework for Energy-Efficient and Collision-Free Path Planning in Autonomous Maritime Navigation}

\author{%
\IEEEauthorblockN{\authorfont 1\textsuperscript{st} Yaoze Liu~${\dagger}$}
\IEEEauthorblockA{\affilfont \textit{School of Engineering, University of Strathclyde,}\\
Glasgow, United Kingdom}
\and
\IEEEauthorblockN{\authorfont 2\textsuperscript{nd} Zhen Tian~${\dagger}$}
\IEEEauthorblockA{\affilfont \textit{James Watt School of Engineering, University of Glasgow,}\\
Glasgow, United Kingdom}
\and
\IEEEauthorblockN{\authorfont 3\textsuperscript{rd} Qifan Zhou}
\IEEEauthorblockA{\affilfont \textit{School of Power and energy, Northwestern Polytechnical University,}\\
Shanxi, Xi'an, China}
\and
\IEEEauthorblockN{\authorfont 4\textsuperscript{th} Zixuan Huang}
\IEEEauthorblockA{\affilfont \textit{James Watt School of Engineering, University of Glasgow,}\\
Glasgow, United Kingdom}
\and
\IEEEauthorblockN{\authorfont 5\textsuperscript{th} Hongyu Sun}
\IEEEauthorblockA{\affilfont \textit{School of Engineering, University of Birmingham,}\\
Birmingham, United Kingdom\\
Corresponding author: 2533620@student.bham.ac.uk}
\thanks{\affilfont ${\dagger}$ Equal contribution}
}

\maketitle

\setlength{\parindent}{0.51cm}
\setlength{\parskip}{6pt}
\linespread{0.95}

\renewenvironment{abstract}{
  \noindent\begin{center}\bfseries Abstract\end{center}%
  \begin{quote}\abstractfont
  \setlength{\parindent}{0.48cm}\setlength{\parskip}{10pt}
}{
  \end{quote}
}

\renewenvironment{IEEEkeywords}{
  \noindent\textbf{Keywords---}\keywordfont
  \setlength{\parindent}{0.48cm}\setlength{\parskip}{6pt}
}{
}

\begin{abstract}
This work presents a novel approach to optimizing energy dispatch in autonomous maritime systems by integrating metaheuristic search with model predictive control. Our proposed ACO--MPC framework combines Ant Colony Optimization (ACO) with Model Predictive Control (MPC) to dynamically generate energy-efficient paths in a simulated sea surface environment, where renewable generation is influenced by wind speed and polar effects. A linear model, derived from real-world data, is embedded within the MPC formulation to accurately predict energy consumption, thereby enabling real-time optimization of renewable utilization, battery cycling, and backup power usage. Simulation results demonstrate that the ACO--MPC approach significantly outperforms conventional rule-based strategies and standard MPC methods, achieving both collision-free navigation and the lowest cumulative energy during the navigation to target points.
\end{abstract}


\section{Introduction}
\begin{figure}[t]
\centering
\includegraphics[width=0.9\linewidth]{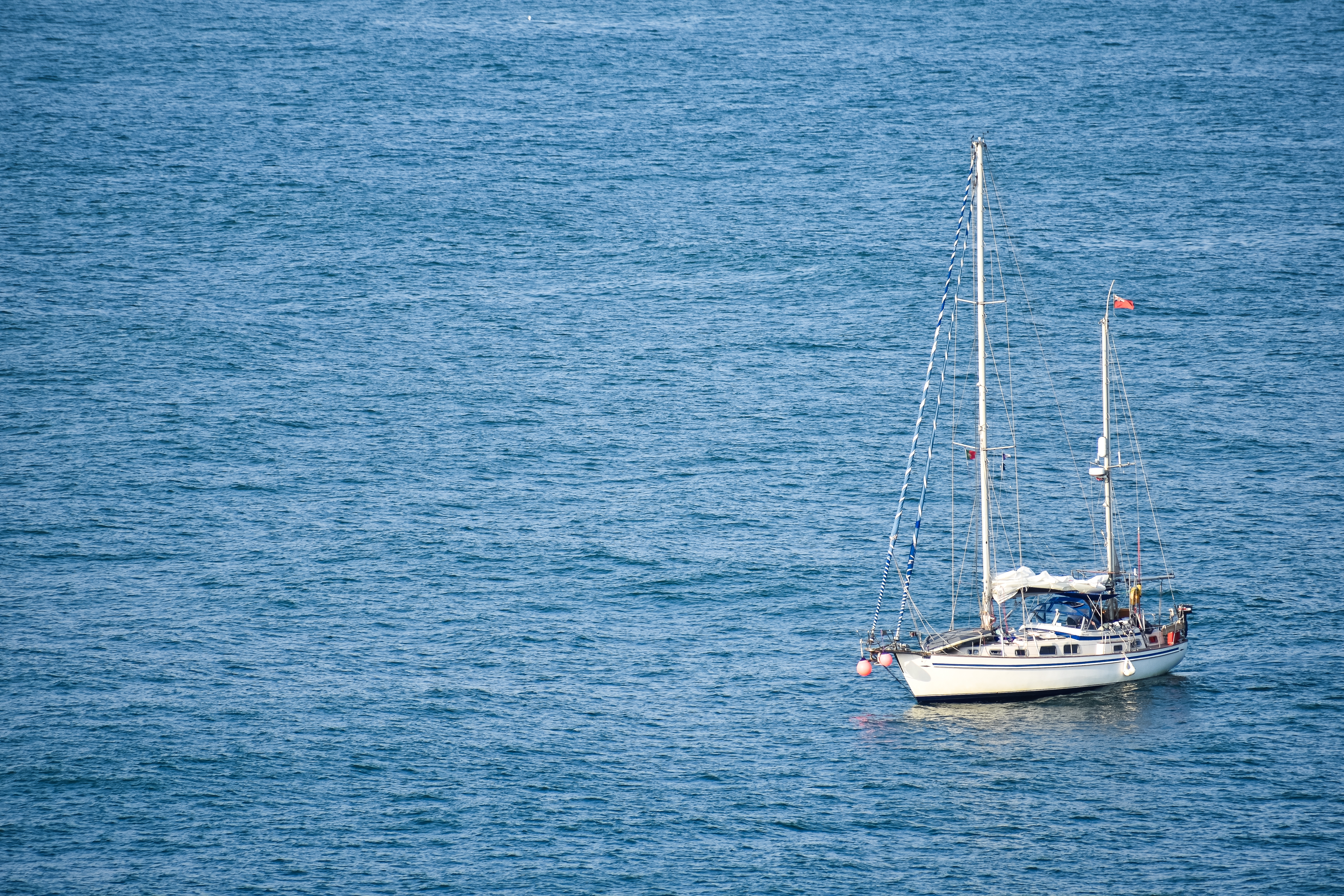}
\vspace{-1mm}
\caption{The example of an autonomous ship.}
\label{fig1_framework}
\end{figure}

Recent advancements in autonomous navigation algorithms have garnered significant attention across various domains~\cite{lin2024enhanced,lin2024conflicts,tian2024balanced,10607945}. Among these, path planning for autonomous ships has emerged as a critical challenge, particularly in complex marine environments with obstacle constraints~\cite{wang2024research,leite2024robocentric,kim2024field,ali2024autonomous}, as illustrated in Fig.~\ref{fig1_framework}. Autonomous ship navigation faces numerous real-world challenges. For instance, offshore aquaculture facilities typically rely on floating platforms that demand a reliable and economical power supply to support operations such as aeration, lighting, monitoring, dead fish removal, and feeding systems~\cite{syse2016investigating,saha2022profit}. In addition, the presence of islets or other obstacles in the sea increases the risk of collisions, which can lead to substantial economic losses~\cite{vatle2024green,atecs2024pompeii}. Therefore, efficient path planning is not only essential for safe navigation but also for reducing energy consumption—a key factor in extending operational range and minimizing environmental impact~\cite{zhang2024review,wang2025computational}.
\begin{figure*}[ht]
    \centering
    \includegraphics[width=1\textwidth]{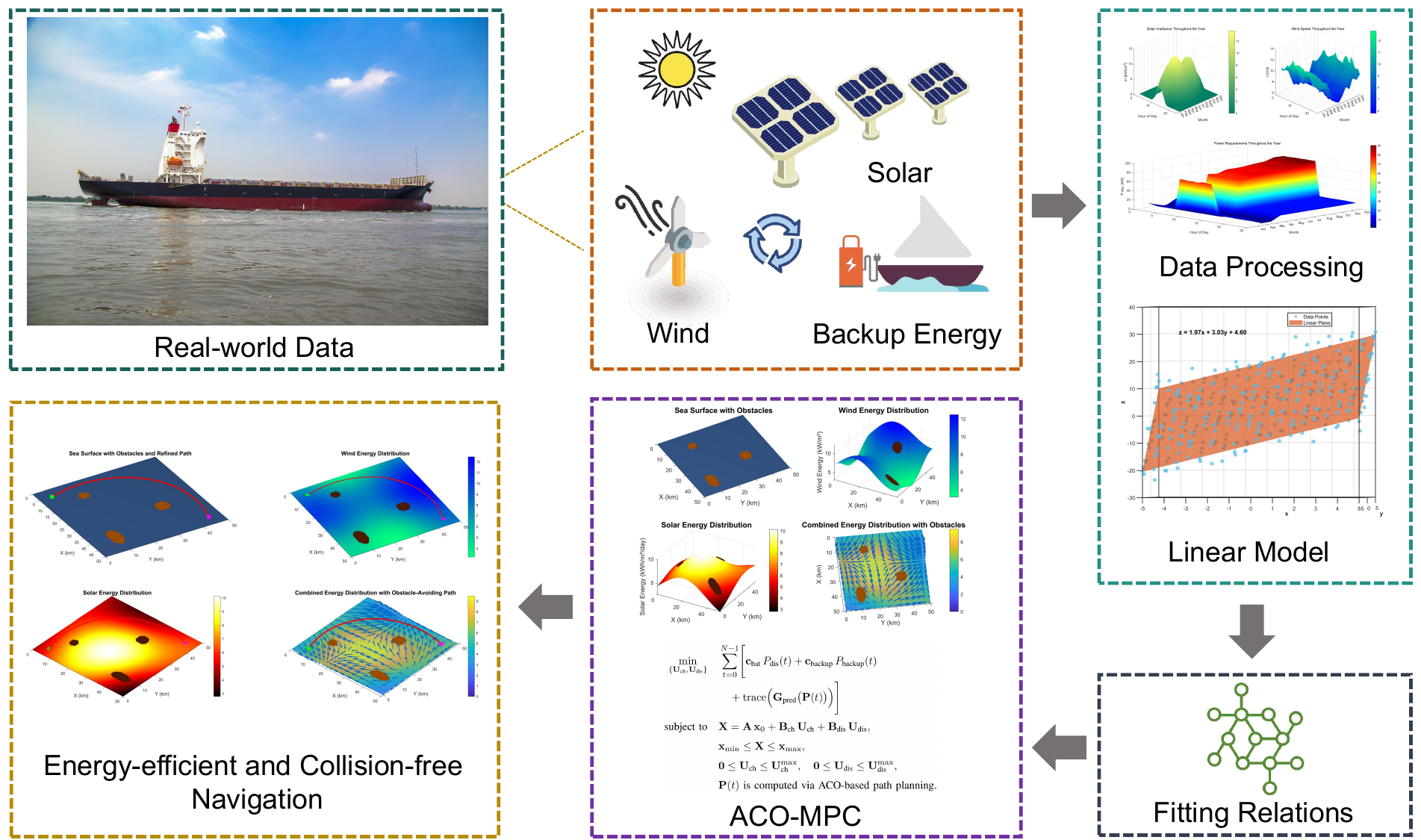}
    \caption{High-level overview of the proposed ACO--MPC framework for autonomous ship navigation. 
    Beginning with real-world data collection, the system preprocesses this information and fits a linear model relating renewable resources to energy consumption. These insights feed into the ACO--MPC module, where path planning decisions are continuously optimized to balance energy efficiency, collision avoidance, and operational constraints. The final output is an energy-efficient and collision-free navigation strategy tailored for complex maritime environments.}
    \label{fig:aco_mpc_framework}
\end{figure*}
Marine navigation is inherently complex due to dynamic sea conditions, including fluctuating wind speeds, varying currents, and unpredictable obstacles~\cite{wu2024review}. Traditional path planning approaches often rely on deterministic algorithms that can struggle with the high energy costs associated with maritime operations. Given that energy consumption in autonomous ships is influenced by multiple environmental factors, such as polar strength and wind patterns, developing energy-efficient path planning strategies is crucial for sustainable maritime operations. To this end, integrating renewable energy sources—such as solar and wind power, in conjunction with battery storage and diesel backup—offers a promising solution~\cite{al2017review,roy2018electrical}.

Model Predictive Control (MPC) has been widely adopted for path planning in autonomous systems because of its ability to handle multi-variable control problems with constraints and predict future states~\cite{kouvaritakis2016model,schwenzer2021review}. Recently, MPC has been applied to energy optimization scenarios~\cite{mesquita2020recent,dokeroglu2019survey}. Complementing MPC, metaheuristic algorithms provide robust methods for efficiently exploring the solution space. In particular, Ant Colony Optimization (ACO) has demonstrated strong performance in path finding by simulating the foraging behavior of ant colonies, where pheromone-based feedback guides the collective search~\cite{dorigo2007ant,wang2024obstacle,heng2024exploring}. When integrated with MPC, ACO enhances the decision-making process by efficiently navigating the trade-offs between energy efficiency and collision avoidance.

This paper introduces an ACO-MPC framework for energy-efficient path planning in autonomous ships operating in sea environments characterized by varying polar strength and wind conditions. Our approach employs pheromone-based learning within a receding-horizon optimization framework to continuously refine ship trajectories, as depicted in Fig.~\ref{fig1_framework}.
First, real-world data are collected to characterize wind resources, solar irradiance, and 
potential backup energy systems. This information is processed and then used to fit a linear model that estimates energy consumption based on environmental parameters. Next, the ACO--MPC module leverages this model to identify 
collision-free paths that minimize overall energy cost, using pheromone-based search and 
receding-horizon optimization. By iterating through these steps in real time, the proposed 
framework ensures safe, robust, and efficient autonomous ship navigation across dynamic sea 
conditions. The key contributions of this work are summarized as follows:
\begin{enumerate}
    \item We propose an energy consumption model that quantifies ship movement costs based on environmental factors, including polar strength and wind speed, using real-world datasets. This model forms the basis for our energy-efficient path planning.
    \item We introduce an ACO-MPC approach that integrates pheromone-based learning with receding-horizon control, enabling the dynamic optimization of ship trajectories while balancing energy efficiency and navigational constraints.
    \item We evaluate the proposed framework under diverse environmental scenarios and compare its performance with conventional MPC and other metaheuristic approaches such as GA-MPC, PSO-MPC, and rule-based methods. Our results demonstrate that the ACO-MPC approach achieves significantly lower energy consumption and avoids collisions, outperforming benchmark algorithms.
\end{enumerate}

The remainder of this paper is organized as follows: Section II presents related work in path planning algorithms for autonomous navigation. Section III describes the problem formulation and the energy consumption model. Section IV details the proposed ACO-MPC algorithm. Section V presents experimental results and comparative analysis. Finally, Section VI concludes the paper and discusses future research directions.

The remainder of this paper is structured as follows. Section 2 presents the ACO-MPC path planning. Section 3 details the linear model fitting. Section 4 discusses the simulation setup and results. Finally, Section 5 provides conclusions and future research directions.
\section{ACO-Based MPC Path Planning}
To ensure that an autonomous ship reaches a specified target with minimal energy loss, we propose an ACO-based Model Predictive Control (MPC) path planning algorithm. In this approach, candidate paths are generated over a prediction horizon \( H \) using matrix formulations to represent the path states. The algorithm integrates an energy loss map \(\mathbf{E}\) and an obstacle matrix \(\mathbf{O}\) to perform a global search for the optimal route from an initial position 
\begin{equation}
\mathbf{x}_0 = \begin{bmatrix} x_0 \\ y_0 \end{bmatrix},
\end{equation}
to a target position 
\begin{equation}
\mathbf{x}_f = \begin{bmatrix} x_f \\ y_f \end{bmatrix}.
\end{equation}

\subsection{System Description and Extended Variable Definitions}
Let the operational domain be discretized into a grid with coordinate matrices \(\mathbf{X} \in \mathbb{R}^{n_y \times n_x}\) and \(\mathbf{Y} \in \mathbb{R}^{n_y \times n_x}\) such that the grid point \((i,j)\) corresponds to the position
\begin{equation}
\mathbf{p}_{i,j} = \begin{bmatrix} x_{i,j} \\ y_{i,j} \end{bmatrix}.
\end{equation}
The energy loss over the grid is represented by the matrix 
\begin{equation}
\mathbf{E} \in \mathbb{R}^{n_y \times n_x},
\end{equation}
where each element \(E_{i,j}\) (in kWh/km) denotes the energy cost associated with cell \((i,j)\). Obstacles are described by the binary matrix 
\begin{equation}
\mathbf{O} \in \{0,1\}^{n_y \times n_x},
\end{equation}
with \(O_{i,j}=1\) indicating an obstacle and \(O_{i,j}=0\) indicating free space.

We denote the global planned path by the matrix
\begin{equation}
\mathbf{P} = \begin{bmatrix}
\mathbf{p}_0^T\\[0.5em]
\mathbf{p}_1^T\\[0.5em]
\vdots\\[0.5em]
\mathbf{p}_K^T
\end{bmatrix}, \quad \text{with} \quad \mathbf{p}_k = \begin{bmatrix} x_k \\ y_k \end{bmatrix},
\end{equation}
where \(K\) is the total number of steps taken.

\subsection{Candidate Path Generation: Matrix-Based Ant Colony Formulation}
Within the prediction horizon \( H \), the possible discrete moves are defined by the matrix
\begin{equation}
\mathbf{M} = \begin{bmatrix}
1 & 0\\[0.5em]
1 & 1\\[0.5em]
0 & 1\\[0.5em]
-1 & 1\\[0.5em]
-1 & 0\\[0.5em]
-1 & -1\\[0.5em]
0 & -1\\[0.5em]
1 & -1
\end{bmatrix} \in \mathbb{R}^{N_m \times 2},
\end{equation}
where \( N_m = 8 \) denotes the number of available movement directions.

\subsubsection{Ant Population Representation}
The candidate path generation is performed by a colony of \( N_a \) ants. We represent the entire ant population’s candidate paths by a three-dimensional matrix
\begin{equation}
\mathcal{P} \in \mathbb{R}^{N_a \times (H+1) \times 2},
\end{equation}
where the \(a\)th candidate path is given by
\begin{equation}
\mathcal{P}(a,:,:) = \mathbf{P}^{(a)} = \begin{bmatrix}
\mathbf{p}_0^{(a)}\\[0.5em]
\mathbf{p}_1^{(a)}\\[0.5em]
\vdots\\[0.5em]
\mathbf{p}_H^{(a)}
\end{bmatrix}, \quad \text{with } \mathbf{p}_0^{(a)} = \mathbf{x}_0.
\end{equation}

\subsubsection{Move Selection Using Matrix Probabilities}
At each prediction step \( h \) (with \( h = 0,1,\dots,H-1 \)), each ant selects a move from the set of possible moves. The selection is based on a probability that fuses both pheromone information and a heuristic measure. The pheromone levels are stored in a matrix
\begin{equation}
\boldsymbol{\Phi} \in \mathbb{R}^{H \times N_m},
\end{equation}
with an initial uniform value:
\begin{equation}
\phi_{h,m} = \phi_0,\quad \forall\, h=1,\dots,H,\; m=1,\dots,N_m.
\end{equation}
For each ant \( a \) and each step \( h \), the heuristic value for move \( m \) is defined as:
\begin{equation}
\eta_{h,m}^{(a)} = \frac{1}{\left\| \mathcal{P}(a,h,:) + \mathbf{M}(m,:) - \mathbf{x}_f^T \right\|_2 + \epsilon},
\end{equation}
where \(\epsilon > 0\) prevents division by zero. The probability that an ant \(a\) selects move \( m \) at step \( h \) is then given by
\begin{equation}
p_{h,m}^{(a)} = \frac{\phi_{h,m} \cdot \eta_{h,m}^{(a)}}{\sum_{j=1}^{N_m} \phi_{h,j} \cdot \eta_{h,j}^{(a)}}.
\end{equation}
In matrix form, one can view the probabilities for all moves at step \( h \) for ant \( a \) as the vector
\begin{equation}
\mathbf{p}_h^{(a)} = \begin{bmatrix} p_{h,1}^{(a)} \\[0.5em] p_{h,2}^{(a)} \\[0.5em] \vdots \\[0.5em] p_{h,N_m}^{(a)} \end{bmatrix}.
\end{equation}

\subsubsection{Validity and Update of Candidate Paths}
To ensure feasibility, a validity function \( \mathcal{V}: \mathbb{R}^2 \to \{0,1\} \) is defined as:
\begin{equation}
\mathcal{V}(\mathbf{p}) = 
\begin{cases}
1, & \text{if } \mathbf{p} \text{ is within the grid bounds and } \mathbf{O}(p_y,p_x)=0, \\[1mm]
0, & \text{otherwise.}
\end{cases}
\end{equation}
At each step \( h \), if an ant \( a \) selects move \( m^* \) (according to the probability vector \(\mathbf{p}_h^{(a)}\)), its next position is updated as
\begin{equation}
\mathcal{P}(a,h+1,:) = \mathcal{P}(a,h,:) + \mathbf{M}(m^*,:)^T.
\end{equation}
If the resulting position does not satisfy \( \mathcal{V}(\cdot)=1 \), the candidate path is marked as invalid (and its cost is set to infinity).

\subsubsection{Extended Candidate Cost Evaluation}
For each candidate path \( \mathbf{P}^{(a)} \), the cumulative cost is defined as:
\begin{equation}
J\left(\mathbf{P}^{(a)}\right) = \sum_{i=1}^{H} \left[ \omega_i \cdot E\left( p_{i,1}^{(a)},\, p_{i,2}^{(a)} \right) + \lambda \cdot \mathcal{I}\Bigl( \mathcal{V}\left( \mathbf{p}_i^{(a)} \right) = 0 \Bigr) \right],
\end{equation}
where:
\begin{itemize}
    \item \( \omega_i \) are stage-dependent weight factors (often set uniformly, i.e., \( \omega_i = 1 \)),
    \item \( \lambda \gg 0 \) is a penalty weight for infeasible moves,
    \item \( \mathcal{I}(\cdot) \) is the indicator function.
\end{itemize}
In matrix notation, the cost vector for candidate \( a \) can be written as
\begin{equation}
\begin{split}
\mathbf{c}^{(a)} &= \begin{bmatrix} 
c_1^{(a)} \\[0.5em] 
c_2^{(a)} \\[0.5em] 
\vdots \\[0.5em] 
c_H^{(a)} 
\end{bmatrix}, \quad \text{with} \\
c_i^{(a)} &= \omega_i \cdot E\Bigl( p_{i,1}^{(a)},\, p_{i,2}^{(a)} \Bigr) \\
&\quad + \lambda \cdot \mathcal{I}\Bigl( \mathcal{V}\bigl( \mathbf{p}_i^{(a)} \bigr)=0 \Bigr).
\end{split}
\end{equation}

Then the total cost is given by
\begin{equation}
J\left(\mathbf{P}^{(a)}\right) = \mathbf{1}^T \mathbf{c}^{(a)},
\end{equation}
where \(\mathbf{1}\) is an \( H \times 1 \) vector of ones.

\subsubsection{Pheromone Update in Matrix Form}
After all \( N_a \) candidate paths are generated, the pheromone matrix is updated to reinforce successful moves. For each prediction step \( h \) and move \( m \), the pheromone update rule is:
\begin{equation}
\phi_{h,m} \leftarrow (1-\rho)\,\phi_{h,m} + \sum_{a=1}^{N_a} \delta_{h,m}^{(a)},
\end{equation}
where the evaporation rate is \( \rho \in (0,1) \) and the reinforcement term is defined as
\begin{equation}
\delta_{h,m}^{(a)} = 
\begin{cases}
\displaystyle \frac{1}{J\left(\mathbf{P}^{(a)}\right)}, & \text{if candidate } a \text{ used move } m \text{ at step } h, \\[2mm]
0, & \text{otherwise.}
\end{cases}
\end{equation}
In a fully matrix-based approach, one may define an incidence matrix 
\begin{equation}
\mathbf{I}^{(a)} \in \{0,1\}^{H \times N_m},
\end{equation}
for each candidate path, where the \((h,m)\)th element is 1 if ant \(a\) took move \(m\) at step \(h\) and 0 otherwise. Then the pheromone update can be compactly written as:
\begin{equation}
\boldsymbol{\Phi} \leftarrow (1-\rho)\,\boldsymbol{\Phi} + \sum_{a=1}^{N_a} \frac{\mathbf{I}^{(a)}}{J\left(\mathbf{P}^{(a)}\right)}.
\end{equation}

The complete ACO--MPC framework is therefore governed by the following matrix equations:
\begin{enumerate}[label=\arabic*.]
    \item \textbf{Initialization:} Set the candidate path population:
    \begin{equation}
    \mathcal{P}(a,1,:) = \mathbf{x}_0, \quad \forall\, a=1,\dots,N_a,
    \end{equation}
    and initialize \(\boldsymbol{\Phi}\) as:
    \begin{equation}
    \phi_{h,m} = \phi_0, \quad \forall\, h=1,\dots,H,\; m=1,\dots,N_m.
    \end{equation}
    
    \item \textbf{Move Selection:} For each ant \(a\) and each step \(h\):
    \begin{equation}
    p_{h,m}^{(a)} = \frac{\phi_{h,m} \cdot \eta_{h,m}^{(a)}}{\sum_{j=1}^{N_m} \phi_{h,j} \cdot \eta_{h,j}^{(a)}},
    \end{equation}
    with
    \begin{equation}
    \eta_{h,m}^{(a)} = \frac{1}{\left\|\mathcal{P}(a,h,:) + \mathbf{M}(m,:) - \mathbf{x}_f^T\right\|_2 + \epsilon}.
    \end{equation}
    
    \item \textbf{Path Update:} Update candidate paths by:
    \begin{equation}
    \mathcal{P}(a, h+1,:) = \mathcal{P}(a, h, :) + \mathbf{M}(m^*,:)^T,
    \end{equation}
    where \( m^* \) is chosen according to \( \{ p_{h,m}^{(a)} \}_{m=1}^{N_m} \) and subject to \(\mathcal{V}=1\).
    
    \item \textbf{Cost Evaluation:} Compute the cost vector for each candidate:
    \begin{equation}
    \mathbf{c}^{(a)} = \begin{bmatrix} \omega_1 \, E\left(p_{1,1}^{(a)}, p_{1,2}^{(a)}\right) + \lambda \, \mathcal{I}\Bigl(\mathcal{V}(\mathbf{p}_1^{(a)})=0\Bigr) \\[0.5em] \vdots \\[0.5em] \omega_H \, E\left(p_{H,1}^{(a)}, p_{H,2}^{(a)}\right) + \lambda \, \mathcal{I}\Bigl(\mathcal{V}(\mathbf{p}_H^{(a)})=0\Bigr) \end{bmatrix},
    \end{equation}
    and total cost:
    \begin{equation}
    J\left(\mathbf{P}^{(a)}\right) = \mathbf{1}^T \mathbf{c}^{(a)}.
    \end{equation}
    
    \item \textbf{Pheromone Update:} Update the pheromone matrix as:
    \begin{equation}
    \phi_{h,m} \leftarrow (1-\rho)\phi_{h,m} + \sum_{a=1}^{N_a} \frac{I_{h,m}^{(a)}}{J\left(\mathbf{P}^{(a)}\right)}.
    \end{equation}
\end{enumerate}

This detailed matrix formulation encapsulates the entire ant colony optimization mechanism within the MPC framework, and forms the foundation for the algorithm presented in Algorithm~\ref{alg:ACO_MPC}. The matrices \(\mathcal{P}\), \(\boldsymbol{\Phi}\), and \(\mathbf{I}^{(a)}\) collectively represent the evolution of the ant population and guide the search for the optimal path with minimum energy loss. The pseudocode in Algorithm~\ref{alg:ACO_MPC} (provided previously) integrates these matrix-based formulations to perform iterative path planning. All operations—from candidate path generation and cost evaluation to the matrix-based pheromone update—are carried out using the equations described above.

\begin{algorithm}[t]
\caption{ACO--MPC Path Planning Algorithm}
\label{alg:ACO_MPC}
\begin{algorithmic}[1]
    \State \textbf{Input:} Initial state \(\mathbf{x}_0\), target state \(\mathbf{x}_f\), energy loss map \(\mathbf{E}\), obstacle matrix \(\mathbf{O}\), prediction horizon \( H \), maximum iterations \( I_{\max} \).
    \State Initialize global path: \(\mathbf{P} \leftarrow \mathbf{x}_0\); set current state \(\mathbf{x} \leftarrow \mathbf{x}_0\).
    \For{\( i = 1 \) to \( I_{\max} \)}
        \If{\(\|\mathbf{x} - \mathbf{x}_f\|_2 < \epsilon\)}
            \State \(\mathbf{P} \leftarrow [\mathbf{P}; \mathbf{x}_f]\);
            \State \textbf{break};
        \EndIf
        \State Initialize pheromone matrix \(\boldsymbol{\Phi} \in \mathbb{R}^{H \times N_m}\) with \(\phi_0\).
        \For{\( g = 1 \) to \( G_{\text{ACO}} \)}
            \For{each ant \( a = 1,\dots, N_a \)}
                \State Set candidate path: \(\mathcal{P}(a,1,:) \leftarrow \mathbf{x}\);
                \For{\( h = 1 \) to \( H \)}
                    \State Compute move probabilities \( p_{h,m}^{(a)} \) using:
                    \[
                    p_{h,m}^{(a)} = \frac{\phi_{h,m} \cdot \left(\frac{1}{\|\mathcal{P}(a,h,:) + \mathbf{M}(m,:) - \mathbf{x}_f^T\|_2 + \epsilon}\right)}{\sum_{j=1}^{N_m} \phi_{h,j} \cdot \left(\frac{1}{\|\mathcal{P}(a,h,:) + \mathbf{M}(j,:) - \mathbf{x}_f^T\|_2 + \epsilon}\right)},
                    \]
                    where only valid moves (i.e., those satisfying \(\mathcal{V} = 1\)) are considered.
                    \State Randomly select move \( m^* \) according to \( \{ p_{h,m}^{(a)} \}_{m=1}^{N_m} \);
                    \State Update candidate path:
                    \[
                    \mathcal{P}(a,h+1,:) \leftarrow \mathcal{P}(a,h,:) + \mathbf{M}(m^*,:)^T.
                    \]
                \EndFor
                \State Compute cost \( J\left(\mathcal{P}(a,:,:)\right) = \sum_{j=1}^{H} \omega_j \cdot \mathbf{E}\bigl(\mathcal{P}(a,j,1),\,\mathcal{P}(a,j,2)\bigr) \);
            \EndFor
            \State Update pheromone matrix \(\boldsymbol{\Phi}\) as:
            \[
            \phi_{h,m} \leftarrow (1-\rho)\,\phi_{h,m} + \sum_{a=1}^{N_a} \delta_{h,m}^{(a)},
            \]
            with
            \[
            \delta_{h,m}^{(a)} = 
            \begin{cases}
            \dfrac{1}{J\left(\mathcal{P}(a,:,:)\right)}, & \text{if ant } a \text{ used move } m \text{ at step } h, \\[1mm]
            0, & \text{otherwise.}
            \end{cases}
            \]
        \EndFor
        \State Select candidate path with minimum cost, i.e., 
        \(\mathcal{P}^* = \arg \min_{a}\, J\left(\mathcal{P}(a,:,:)\right)\);
        \State Update current state: \(\mathbf{x} \leftarrow \mathcal{P}^*(2,:)\);
        \State Append current state: \(\mathbf{P} \leftarrow [\mathbf{P}; \mathbf{x}]\);
    \EndFor
    \State \textbf{Output:} Global path matrix \(\mathbf{P}\).
\end{algorithmic}
\end{algorithm}

\section{Linear Model Fitting for Energy Consumption and Matrix-Based MPC Formulation}

To further improve the overall control strategy, we integrate a linear model to fit the energy consumption and then embed this model within a matrix-based MPC formulation. In this approach, the energy consumption per unit distance (in kWh/km) is approximated as a linear function of key environmental factors. For instance, let the estimated energy consumption be modeled by
\begin{equation}
G = \gamma_1 \, \mathbf{R}_{\text{pol}} + \gamma_2 \, \mathbf{V}_{\text{wind}} + \gamma_3 \, \mathbf{V}_{\text{wind}}^{(3)} + \gamma_4,
\end{equation}
where \(\mathbf{R}_{\text{pol}}\) is the polar strength map, \(\mathbf{V}_{\text{wind}}\) is the wind speed map, and \(\mathbf{V}_{\text{wind}}^{(3)}\) denotes the element-wise cube of the wind speed. The coefficients \(\gamma_1\), \(\gamma_2\), \(\gamma_3\), and \(\gamma_4\) are determined via least-squares regression. In matrix notation, the predicted energy consumption map is expressed as
\begin{equation}
\mathbf{G}_{\text{pred}} = \gamma_1 \, \mathbf{R}_{\text{pol}} 
+ \gamma_2 \, \mathbf{V}_{\text{wind}} 
+ \gamma_3 \, \mathbf{V}_{\text{wind}}^{(3)} 
+ \gamma_4 \, \mathbf{1},
\end{equation}
where \(\mathbf{1}\) is a matrix of ones matching the dimensions of \(\mathbf{G}_{\text{pred}}\).
\begin{figure*}[t]
    \centering
    \includegraphics[width=0.8\linewidth]{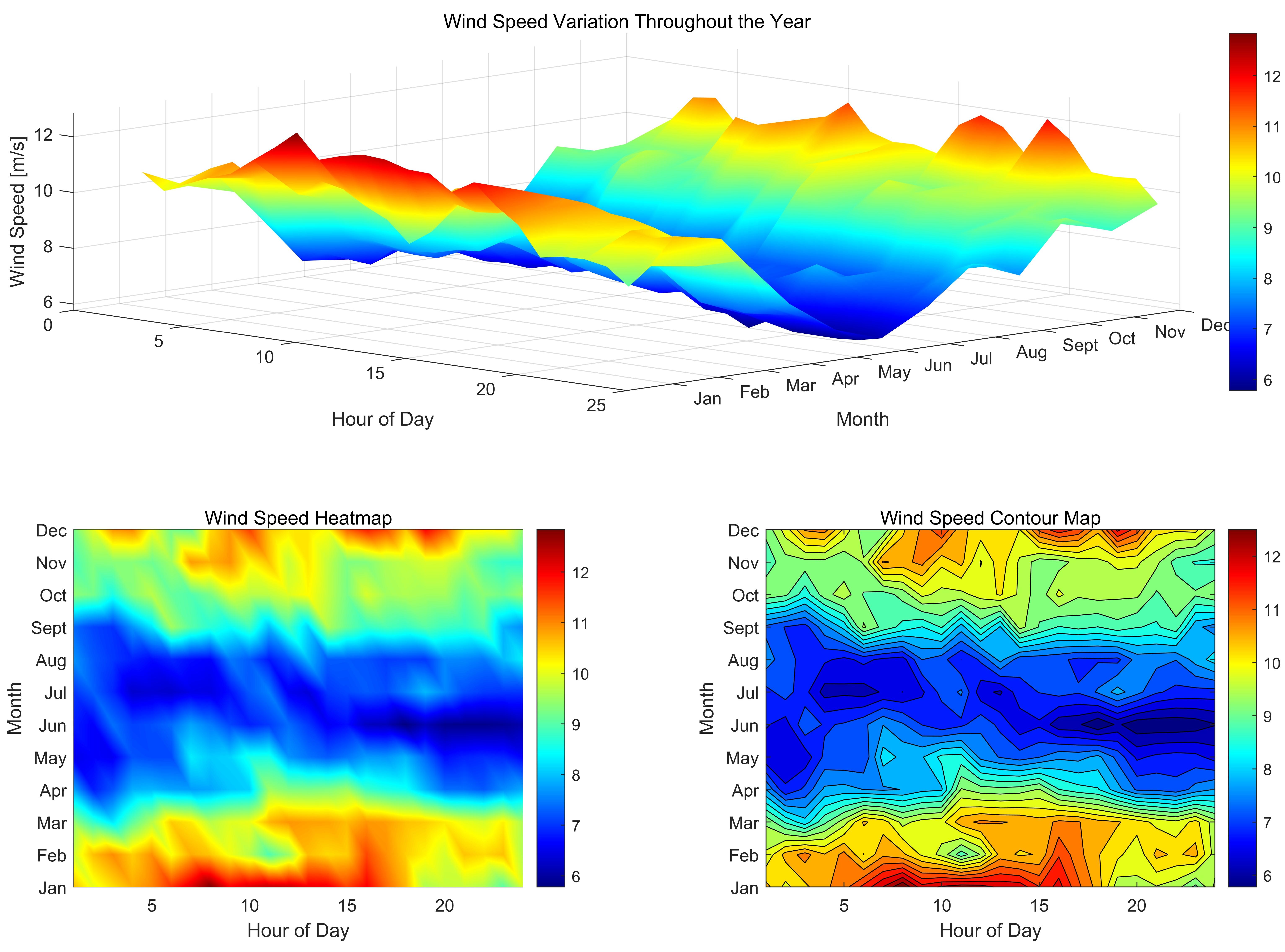}
    \vspace{-2mm}
\caption{The hourly velocity of the ship over 12 months based on HOMER-generated synthetic resource datasets.}
    \label{fig3_lr}
\end{figure*}
This fitted energy model is then integrated into a matrix-based MPC formulation. Let the control vector at time step \(t\) be represented as
\begin{equation}
\mathbf{u}(t) = \begin{bmatrix} P_{\text{ch}}(t) \\ P_{\text{dis}}(t) \end{bmatrix},
\end{equation}
and the state vector (e.g., battery state-of-charge, SOC) be
\begin{equation}
\mathbf{x}(t) = \text{SOC}(t).
\end{equation}
The battery dynamics over a prediction horizon \(N\) are written in matrix form as
\begin{equation}
\mathbf{X} = \mathbf{A} \, \mathbf{x}_0 + \mathbf{B}_{\text{ch}} \, \mathbf{U}_{\text{ch}} + \mathbf{B}_{\text{dis}} \, \mathbf{U}_{\text{dis}},
\end{equation}
where \(\mathbf{X} \in \mathbb{R}^{N \times 1}\) is the predicted SOC trajectory, \(\mathbf{x}_0\) is the initial SOC, and \(\mathbf{A} \in \mathbb{R}^{N \times N}\) is the state transition matrix (often chosen as an identity matrix). The matrices \(\mathbf{B}_{\text{ch}}\) and \(\mathbf{B}_{\text{dis}}\) incorporate charging and discharging efficiencies, respectively, while \(\mathbf{U}_{\text{ch}}\) and \(\mathbf{U}_{\text{dis}}\) are the control input sequences over the horizon.

The cost function to be minimized over the horizon includes both the energy cost derived from the fitted model and additional operational costs. In matrix form, the cumulative cost is given by
\begin{equation}
J = \sum_{t=0}^{N-1} \Bigl[ \mathbf{c}_{\text{bat}} \, P_{\text{dis}}(t) 
+ \mathbf{c}_{\text{backup}} \, P_{\text{backup}}(t) 
+ \mathrm{trace}\Bigl( \mathbf{G}_{\text{pred}}\bigl(\mathbf{P}(t)\bigr) \Bigr) \Bigr],
\end{equation}
where the term \(\mathrm{trace}\Bigl( \mathbf{G}_{\text{pred}}(\mathbf{P}(t)) \Bigr)\) sums the predicted energy consumption along the planned path, with \(\mathbf{P}(t)\) representing the path state at time \(t\).

The MPC optimization problem is formulated as
\begin{equation}
\begin{split}
\min_{\{\mathbf{U}_{\text{ch}},\mathbf{U}_{\text{dis}}\}} \quad & \sum_{t=0}^{N-1} \Biggl[ 
\mathbf{c}_{\text{bat}} \, P_{\text{dis}}(t) + \mathbf{c}_{\text{backup}} \, P_{\text{backup}}(t) \\
& \quad + \mathrm{trace}\Bigl( \mathbf{G}_{\text{pred}}\bigl(\mathbf{P}(t)\bigr) \Bigr) 
\Biggr] \\[1mm]
\text{subject to} \quad 
& \mathbf{X} = \mathbf{A} \, \mathbf{x}_0 + \mathbf{B}_{\text{ch}} \, \mathbf{U}_{\text{ch}} + \mathbf{B}_{\text{dis}} \, \mathbf{U}_{\text{dis}}, \\[1mm]
& \mathbf{x}_{\min} \le \mathbf{X} \le \mathbf{x}_{\max}, \\[1mm]
& \mathbf{0} \le \mathbf{U}_{\text{ch}} \le \mathbf{U}_{\text{ch}}^{\max}, 
\quad \mathbf{0} \le \mathbf{U}_{\text{dis}} \le \mathbf{U}_{\text{dis}}^{\max}, \\[1mm]
& \mathbf{P}(t) \text{ is computed via ACO-based path planning.}
\end{split}
\end{equation}
\begin{figure*}[ht]
    \centering
    \includegraphics[width=0.9\linewidth]{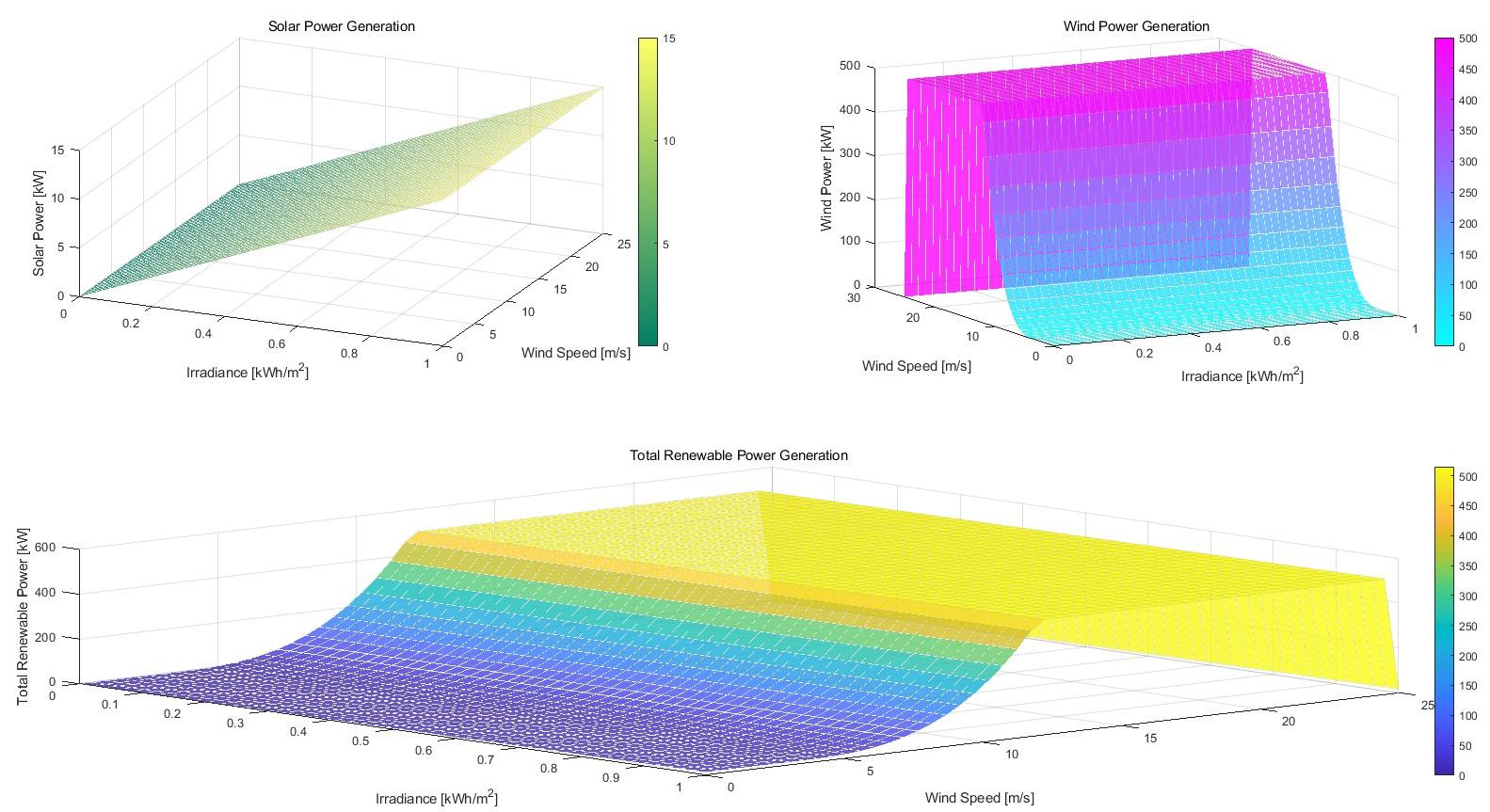}
    \caption{Daily renewable generation profile in June, illustrating solar, wind, and total renewable power.}
    \label{fig:renewable_june}
\end{figure*}
The combined ACO--MPC framework operates as follows. First, ACO is used to generate candidate paths \(\mathbf{P}\) over a short prediction horizon \(H\) that minimize the cumulative energy cost, as evaluated using \(\mathbf{G}_{\text{pred}}\). The candidate path with the lowest energy cost is then selected, and the corresponding control inputs \(\mathbf{U}_{\text{ch}}\) and \(\mathbf{U}_{\text{dis}}\) are determined by solving the MPC optimization problem. Finally, the battery dynamics are updated according to the matrix equation, and the process is repeated in a receding-horizon manner. All operations, including candidate path generation, cost evaluation, and control optimization, are expressed in matrix form to enable efficient computation and real-time implementation.

\section{Simulation Results}
To verify the efficiency of the proposed EG-MPC, simulations are designed. This section presents our simulation results and an
analysis of the proposed framework in Matlab 2024b. In the first subsection, the linear fitting process based on HOMER-generated synthetic resource datasets is illustrated. In the second subsection, the total energy consumption using EG-MPC compared to other popular benchmarks are illustrated. The parameters used are summarized in Table. 1.
\begin{table}[h!]
\centering
\caption{Simulation Parameters and Battery Specifications}
\label{tab:parameters}
\begin{tabular}{|c|c|c|}
\hline
\textbf{Parameter} & \textbf{Value} & \textbf{Description} \\ \hline
\(T_{\text{sim}}\) & 24 & Simulation horizon (hours) \\ \hline
\(\Delta t\) & 1 & Time step (hours) \\ \hline
\(C_{\text{bat}}\) & 1000 & Battery capacity (kWh) \\ \hline
\(\text{SoC}_0\) & 500 & Initial state-of-charge (kWh) \\ \hline
\(P_{\text{ch}}^{\max}\) & 1000 & Maximum charging rate (kW) \\ \hline
\(P_{\text{dis}}^{\max}\) & 100 & Maximum discharging rate (kW) \\ \hline
\(\eta_{\text{bat}}\) & 0.9 & Charging/discharging efficiency \\ \hline
\(v(t)\) & 8 & Fixed wind speed profile (m/s) \\ \hline
\end{tabular}
\end{table}

\begin{figure}[ht]
    \centering
    \includegraphics[width=1\linewidth]{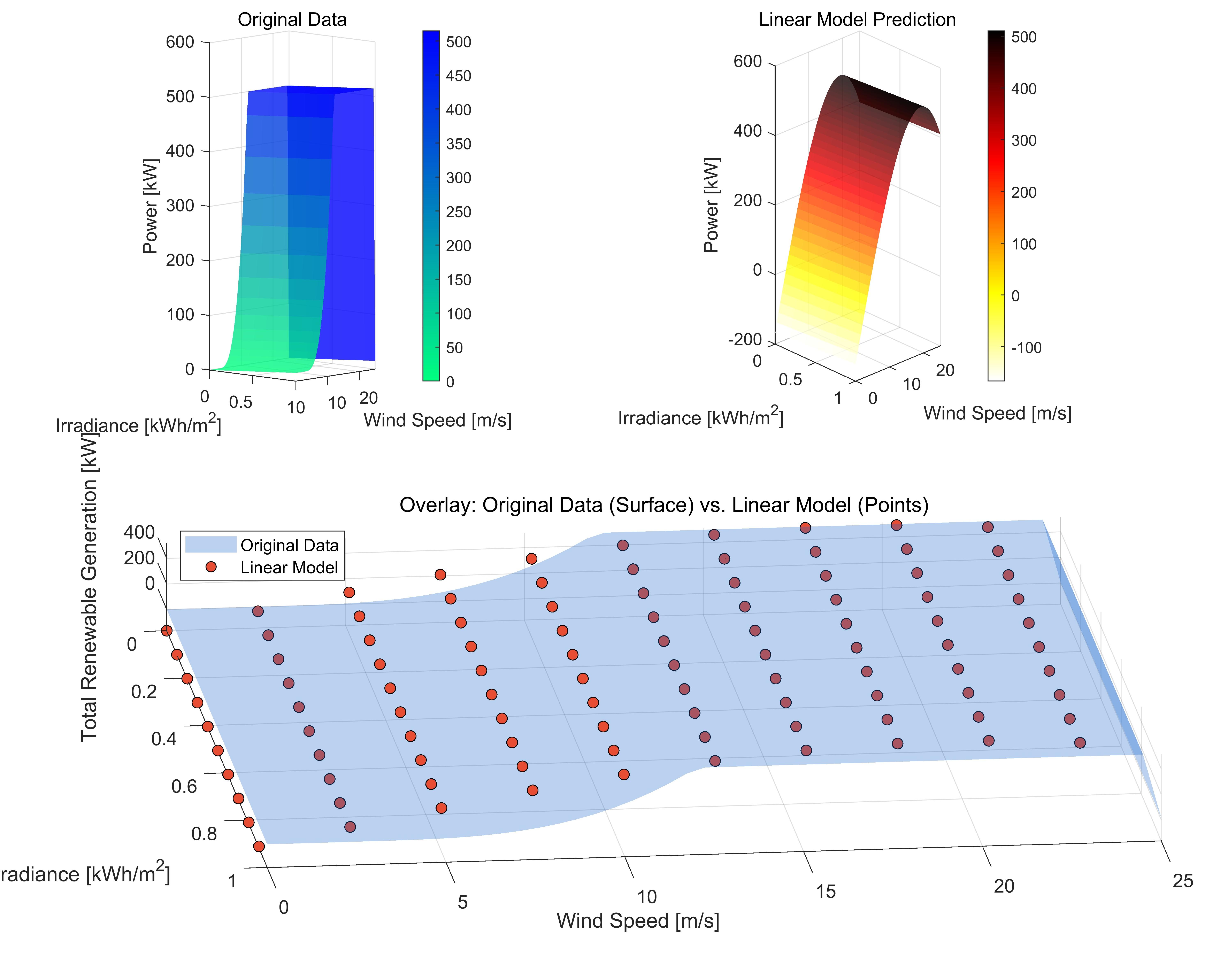}
    \caption{Comparison of original data versus linear model predictions for total renewable generation.}
    \label{fig:comparison_scatter_surface}
\end{figure}
\begin{figure*}[ht]
    \centering
    \includegraphics[width=0.9\linewidth]{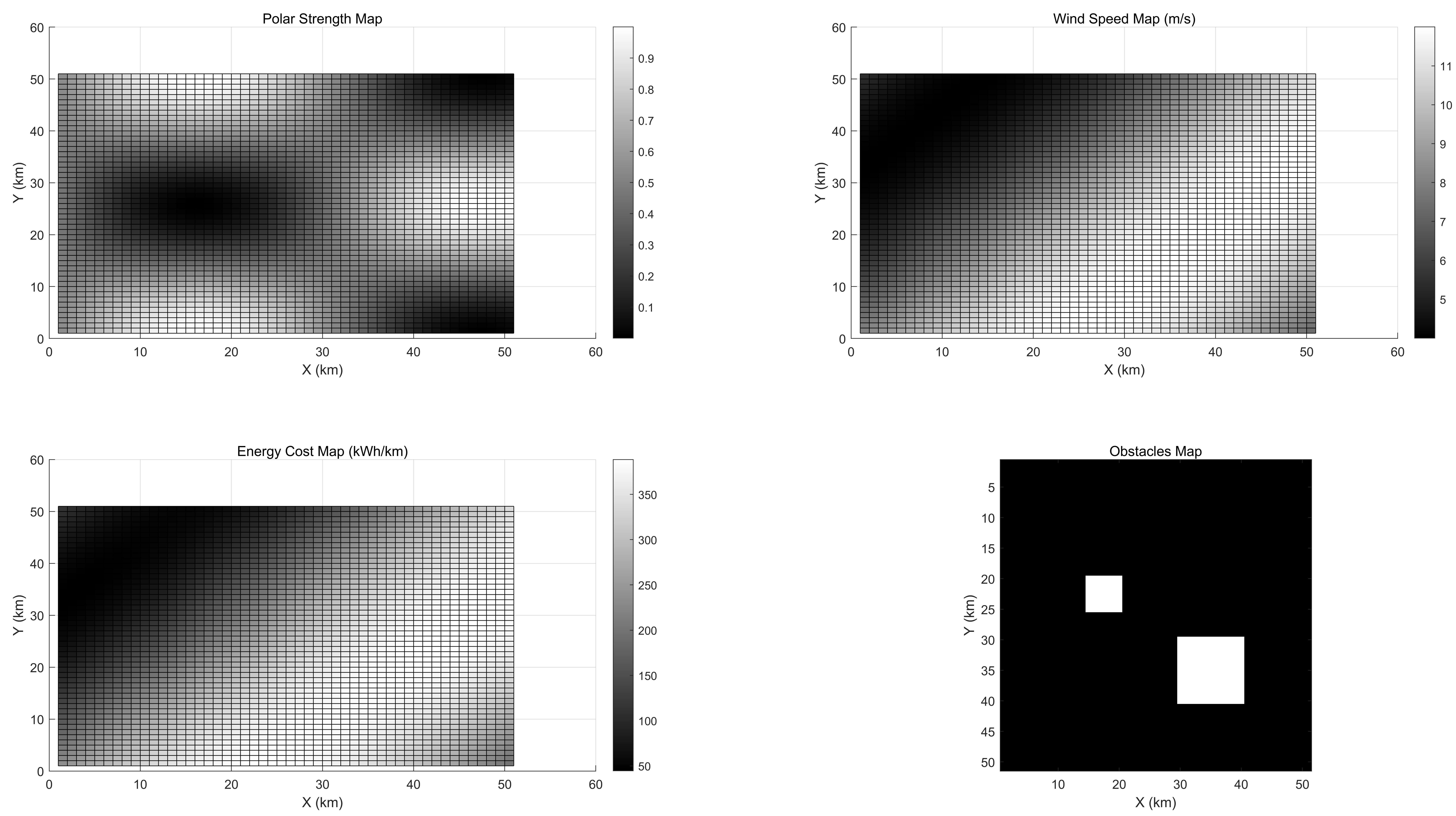}
    \caption{Simulated maritime environment for testing the ACO--MPC path planner.}
    \label{fig:sea_environment}
\end{figure*}
\begin{figure*}[ht]
    \centering
    \includegraphics[width=0.9\linewidth]{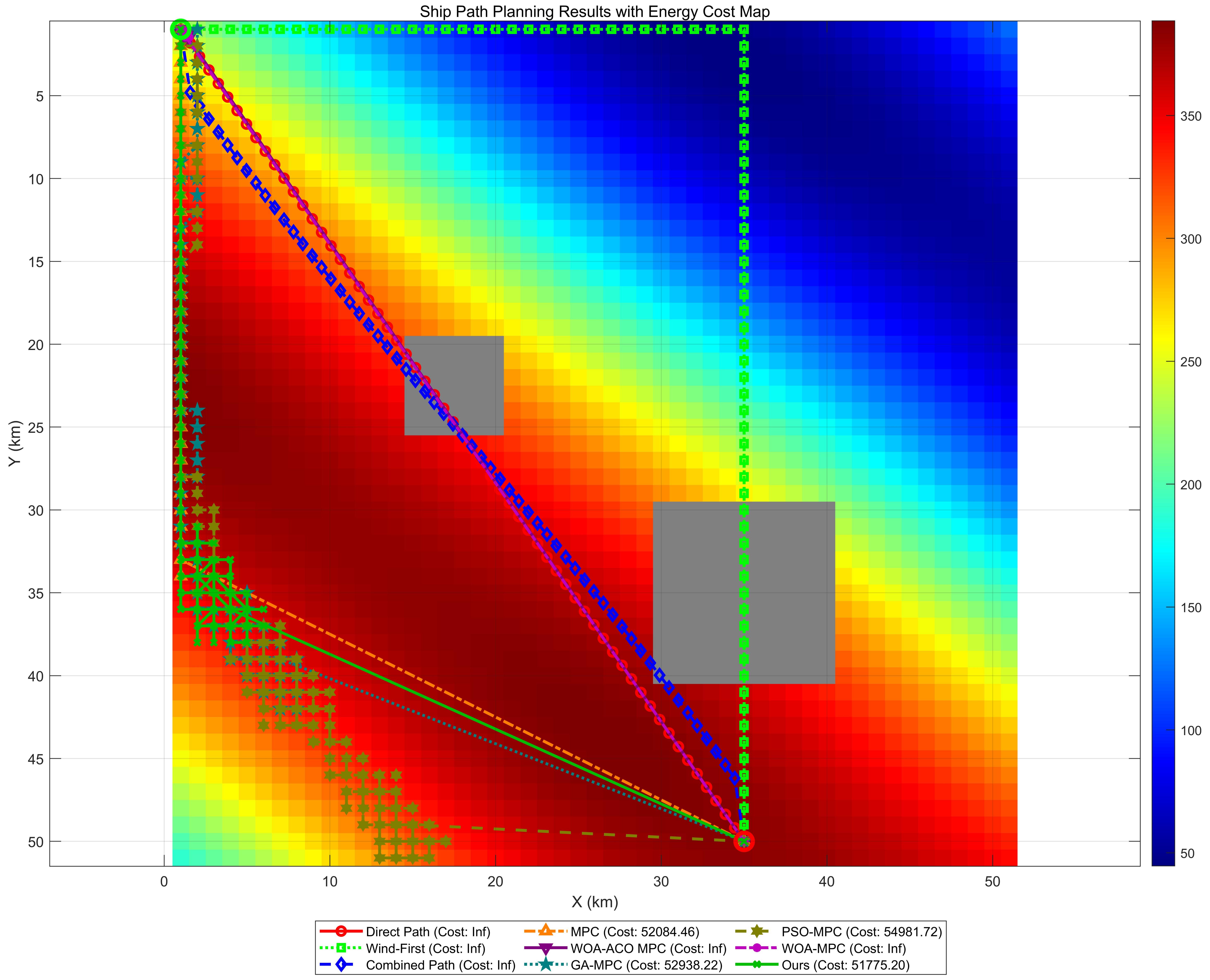}
    \caption{Comparison of path planning algorithms over an energy cost map, with two central square obstacles. 
    Any path intersecting an obstacle is deemed invalid (cost $=\infty$). According to the legend, the total energy 
    cost for each method is as follows: the direct path, wind-first, combined path, WOCA-ACO MPC, WOA-MPC collide with obstacles, resulting in infinite cost; 
    standard MPC consumes 52.084\,kWh; GA-MPC reaches 52.938\,kWh; PSO--MPC uses 54.981\,kWh; and our proposed approach achieves the lowest feasible cost at 51.775\,kWh. This outcome highlights the effectiveness of our method in balancing route length, obstacle avoidance, and energy expenditure, surpassing alternative strategies in a complex maritime environment.}
    \label{fig:path_planning_results1}
\end{figure*}
\begin{figure*}[ht]
    \centering
    \includegraphics[width=0.85\linewidth]{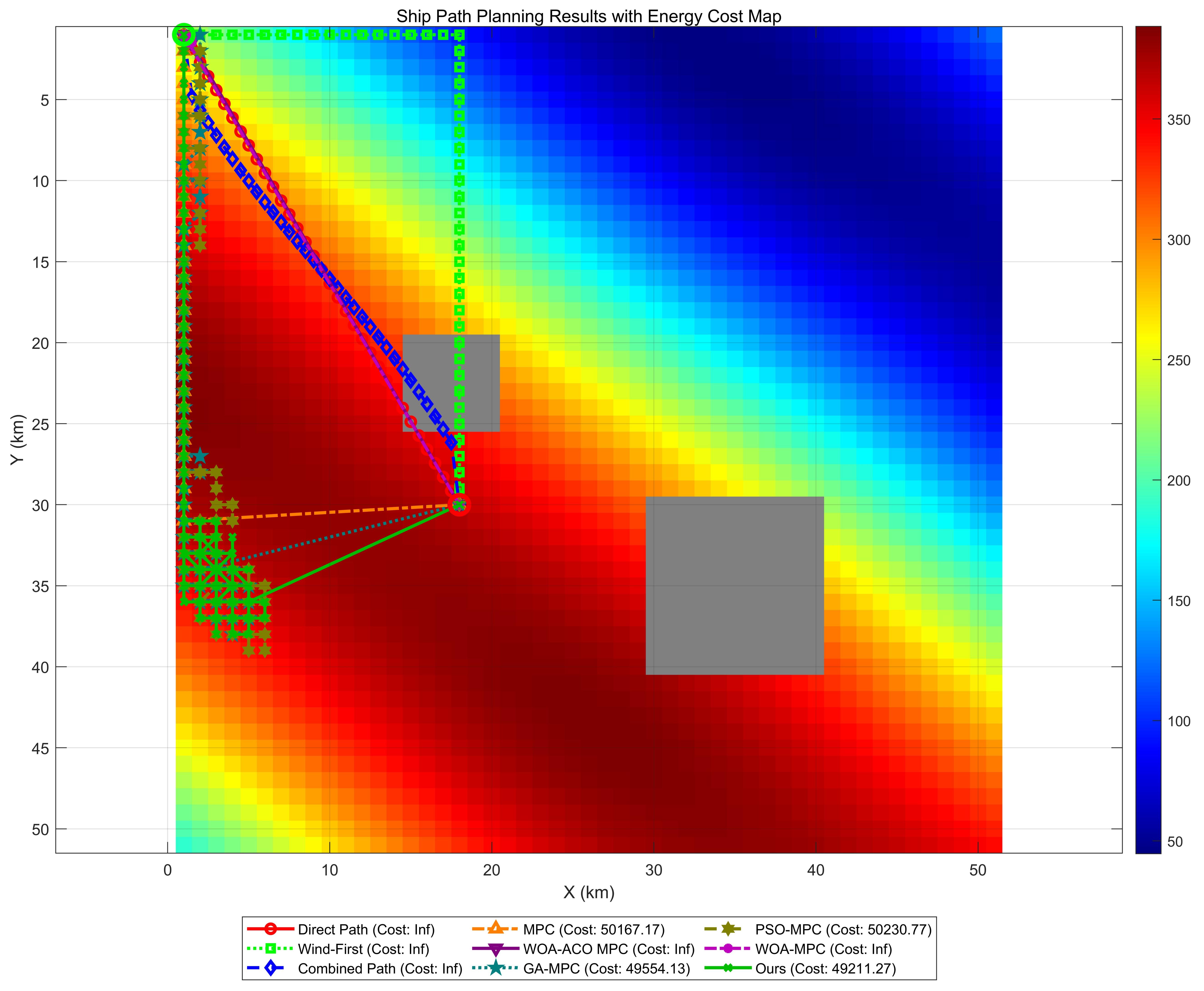}
    \caption{Another Comparison of path planning algorithms over an energy cost map, with two central square obstacles. 
    Any path intersecting an obstacle is deemed invalid (cost $=\infty$). According to the legend, the total energy 
    cost for each method is as follows: the direct path, wind-first, combined path, WOCA-ACO MPC, WOA-MPC collide with obstacles, resulting in infinite cost; 
    standard MPC consumes 50.167\,kWh; GA-MPC reaches 49.554\,kWh; PSO--MPC uses 50.23\,kWh; and our proposed approach achieves the lowest feasible cost at 49.221\,kWh. This outcome highlights the effectiveness of our method in balancing route length, obstacle avoidance, and energy expenditure, surpassing alternative strategies in a complex maritime environment.}
    \label{fig:path_planning_results}
\end{figure*}
\subsection{Linear Fitting Process}
Fig.~\ref{fig3_lr} presents three complementary views of wind speed data collected over an entire year. In the top 3D surface plot, the vertical axis indicates wind speed in meters per second, while the horizontal axes correspond to the hour of the day and the progression of months. The color gradient underscores variations in wind speed intensity, highlighting specific times of day or months where wind speed is notably higher. In the bottom-left heatmap, the x-axis represents the hour of day, and the y-axis lists the months. Warmer shades indicate higher wind speeds, whereas cooler hues signify calmer conditions. This view succinctly reveals diurnal and seasonal wind patterns, such as relatively lower speeds in certain months or peak winds during certain hours. bottom-right contour map reuses the same hour-of-day and month layout but displays lines of constant wind speed. Each contour level represents a specific wind speed, allowing the reader to identify more precise transitions between different speed ranges. This format is especially useful for quickly spotting when wind speeds cross critical thresholds in various periods of the year.

Fig.~\ref{fig:renewable_june} presents three separate 3D surfaces that describe how solar irradiance and wind speed jointly influence renewable energy generation. In the left panel, solar power generation rises in response to increasing irradiance, while wind speed exerts negligible direct impact on solar output. The middle panel focuses on wind power, which depends more strongly on wind speed than on irradiance. Notably, power increases rapidly once wind speed surpasses the cut-in threshold, stabilizing at rated output beyond a certain point. The right panel integrates both solar and wind contributions to form the total renewable power surface. Here, the overall generation benefits from high values of either irradiance or wind speed, but maximum output occurs when both factors are simultaneously elevated. This highlights the complementary nature of solar and wind sources, where strong solar irradiance and sufficient wind speed together yield higher net renewable production. The resulting 3D visualization underscores the importance of capturing both solar and wind parameters within energy models, particularly when seeking to optimize hybrid renewable systems.

The linear model 
\[
P_{\text{total}} 
= -166.3272 
  + 15 \times \mathit{Irr} 
  + 51.7979 \times v 
  - 0.047 \times v^3
\]
provides an accurate approximation of the relationship between solar irradiance, wind speed, and total renewable power output. As illustrated in Fig.~\ref{fig:comparison_scatter_surface}, the model’s predictions closely track real-world measurements across a range of typical operating conditions, indicating that key environmental variables—namely, irradiance and wind speed—are well captured by the chosen functional form, including the cubic wind-speed term.

Figure~\ref{fig:comparison_scatter_surface} illustrates how the linear model 
approximates the original mesh data for total renewable generation. 
The continuous surface shows the simulated generation obtained by combining 
solar and wind across a range of conditions. 
By contrast, the diamond scatter markers represent the model's predicted output. 
The general alignment of these markers with the surface indicates that the model 
captures the primary trends in renewable generation, although minor deviations 
are visible in some regions. Overall, the figure demonstrates that the fitted linear model, 
including a cubic wind-speed term, provides a reasonable approximation 
to the underlying energy generation relationship.
\subsection{Effectiveness in Energy Cost of ACO-MPC}

This subsection has compared the performance of EG-MPC with several popular benchmarks, including a series of popular benchmarks. 

\begin{itemize}
    \item \textbf{Direct Path (Straight Line):}\\
    This method plots a direct line from the start to the end coordinates. It achieves minimal path distance and simplicity, but it disregards wind or polar effects entirely. Consequently, while it can be computationally efficient, it may incur high energy costs if strong adverse conditions (e.g., headwinds or polar influence) exist along the straight line.

    \item \textbf{Wind-First Path:}\\
    Similar to a “Renewable-First” strategy, this approach prioritizes avoiding headwinds by initially moving in a direction that minimizes wind resistance (e.g., moving horizontally) before progressing vertically toward the target. By reducing the impact of adverse wind conditions, this method can lower energy consumption in wind-dominated regions; however, the resulting longer path may lead to suboptimal routing if other environmental factors become significant.

    \item \textbf{50/50 Combined Path:}\\
    Analogous to a “50/50 Split” strategy in energy management, this method attempts to balance wind and polar effects by incorporating a sinusoidal deviation from the straight-line path. Although this strategy mitigates overreliance on a single factor, its fixed ratio may not dynamically adjust to real-time fluctuations in environmental conditions, potentially resulting in increased energy costs when one effect predominates.

    \item \textbf{Standard MPC Path Planning:}\\
    Inspired by traditional MPC applications, this approach repeatedly computes a short-horizon optimal route by minimizing a cost function (typically related to energy consumption). Its ability to adapt to changing conditions is a key strength, though its performance may suffer from local optima if the prediction horizon is too short or if computational resources are limited.

    \item \textbf{WOA-ACO Hybrid MPC:}\\
    This hybrid method integrates Whale Optimization Algorithm (WOA) and Ant Colony Optimization (ACO) within an MPC framework. The WOA component explores the search space through spiral encircling strategies, while ACO refines local path segments via pheromone-based selection. Although this combination can produce robust solutions for nonconvex path planning problems, it increases the algorithm's complexity and demands careful parameter tuning.

    \item \textbf{GA-Based MPC:}\\
    Utilizing Genetic Algorithms (GA), this method encodes candidate paths as chromosomes that evolve through selection, crossover, and mutation. By maintaining a diverse population of solutions, GA-based MPC can explore a wide search space and potentially discover near-optimal paths. However, the evolutionary process may lead to excessive battery cycling and longer convergence times, particularly when the problem size increases.

    \item \textbf{PSO-Based MPC:}\\
    In this approach, Particle Swarm Optimization (PSO) treats each candidate path as a particle with associated velocity and position, which are updated based on personal and global best positions. PSO's balance between exploration and exploitation enables effective convergence toward optimal paths, though it may suffer from premature convergence if the swarm diversity is insufficient.

    \item \textbf{WOA-Based MPC:}\\
    This variant employs the Whale Optimization Algorithm alone to iteratively refine the path through mechanisms such as encircling prey and bubble-net attacking. Its strength lies in effectively navigating complex cost landscapes, but it requires careful calibration of parameters like the spiral coefficient, and its performance may be sensitive to the initial conditions.
\end{itemize}

 Figure~\ref{fig:sea_environment} presents four panels that together form the environment in which our ACO--MPC algorithm is evaluated. 
\textbf{(i)}~The top-left Polar Strength Map shows regions with stronger polar influence. 
\textbf{(ii)}~The top-right Wind Speed Map (m/s) reflects varying wind intensities across the sea surface. 
\textbf{(iii)}~The bottom-left Energy Cost Map (kWh/km) combines polar strength and wind speed to estimate the per-kilometer energy expenditure. 
\textbf{(iv)}~Finally, the bottom-right Obstacles Map marks restricted zones that the path planner must avoid. 
By integrating these spatial distributions, the environment allows us to assess the feasibility, efficiency, and adaptability of our ACO--MPC approach under different wind conditions, polar influences, and navigational constraints.
\subsubsection{Comparison Analysis}

Figure~\ref{fig:path_planning_results1} shows how each algorithm navigates the color-coded energy cost map, where 
higher hues denote regions with increased kWh/km. The ship departs from the top-left corner and must reach the bottom-right corner without intersecting the two central obstacles. While simpler methods such as direct path, fail entirely with infinite cost due to collision. Notably, our proposed approach attains the lowest total energy consumption of 51.775 \,kwh demonstrating superior performance in path optimization under adverse environmental conditions.

Figure~\ref{fig:path_planning_results} shows how each algorithm navigates from the top-left corner to the target point in the middle. Our proposed method achieves the lowest feasible cost of 49.221\,kWh. These outcomes underscore the importance of balancing route length, obstacle avoidance, and localized cost variations, with our approach exhibiting superior performance under complex maritime constraints.

\section{Conclusion}
This work addresses the pressing need for efficient path planning in autonomous maritime operations, where high energy consumption and uncertain environmental conditions pose significant challenges. To tackle this, we developed an ACO--MPC framework that synergizes Ant Colony Optimization with Model Predictive Control, leveraging a linear energy cost model derived from real-world data. By dynamically adjusting the route in response to wind speeds, polar strength, and obstacle layouts, our approach significantly reduces overall energy costs compared to rule-based strategies such as Direct, Wind-First, 50/50 Combined and different MPC variants. Simulation results indicate that the proposed ACO--MPC method not only avoids collisions under diverse sea-surface constraints but also achieves lower cumulative energy consumption. These outcomes underscore the advantages of combining metaheuristic search with receding-horizon optimization for robust, adaptive path planning. Building on these findings, future work could explore cooperative routing among multiple autonomous vessels, further refining the linear model and control mechanisms to accommodate interconnected maritime networks with dynamic interactions and shared resources.

\bibliographystyle{IEEEtran}
\bibliography{IEEEabrv,zq_lib}

\end{document}